\documentclass[10pt,twocolumn,letterpaper]{article}

\usepackage[pagenumbers]{cvpr} 

\definecolor{cvprblue}{rgb}{0.21,0.49,0.74}
\usepackage[pagebackref,breaklinks,colorlinks,allcolors=cvprblue]{hyperref}
\usepackage{multirow}
\usepackage{graphicx}
\usepackage{booktabs}
\usepackage{float}
\usepackage{tikz}
\usepackage{colortbl}
\usepackage{amsmath}

\newcommand{\first}{\cellcolor[HTML]{D5F6C0}}
\newcommand{\secon}{\cellcolor[HTML]{F4FDEF}}
\newcommand{\third}{\cellcolor[HTML]{FFFFFF}}

\newcommand{\BlurText}[1]{
  \noindent 
  \begin{tikzpicture}[baseline]
    \draw [gray!60] node[anchor=base, yshift=0.2ex] at (0,0) {#1}; 
    \draw [gray!60] node[anchor=base, yshift=-0.2ex] at (0,0) {#1}; 
    \draw [black] node[anchor=base] at (0,0) {#1}; 
  \end{tikzpicture}
  \hspace*{-0.8em}
}

\newif\ifshowcomments
\showcommentstrue 
\ifshowcomments
    \newcommand{\comment}[1]{\textcolor{olive}{{\em #1}}}
    \newenvironment{multilinecomment}[1]{\begingroup\color{olive}#1}{\endgroup}
        
    \newcommand{\TK}[1]{\textcolor{orange}{{\em {\bf tk:} #1}}}

\else
    \newcommand{\comment}[1]{}
    
    \newcommand{\TK}[1]{}
    
    \newcommand{\oneline}[1]{}
\fi

\definecolor{gold}{rgb}{1.0, 0.84, 0.0}
\definecolor{silver}{rgb}{0.75, 0.75, 0.75}
\definecolor{bronze}{rgb}{0.8, 0.5, 0.2}

\def\paperID{8421} 
\def\confName{CVPR}
\def\confYear{2025}

\title{\vspace{-4mm} \BlurText{Blurred} LiDAR for Sharper 3D: Robust Handheld \\ 3D Scanning with Diffuse LiDAR and RGB}

\newcommand{\superscript}[1]{\ensuremath{^{\textrm{#1}}}}
\author{
   Nikhil Behari\superscript{1}, Aaron Young\superscript{1},
    Siddharth Somasundaram\superscript{1}, \\
    Tzofi Klinghoffer\superscript{1},
    Akshat Dave \superscript{1},
    Ramesh Raskar\superscript{1}\\
    \superscript{1} Massachusetts Institute of Technology \\
    {\tt\small \{nbehari,aryoung,sidsoma,tzofi,ad74,raskar\}@mit.edu}
    \vspace{-12pt} 
}

\begin{document}

\twocolumn[{%
\renewcommand\twocolumn[1][]{#1}%
\maketitle
\begin{center}
    \centering
    \captionsetup{type=figure}
    \label{fig:teaser}
    \vspace{-4mm}
    \includegraphics[width=\textwidth]{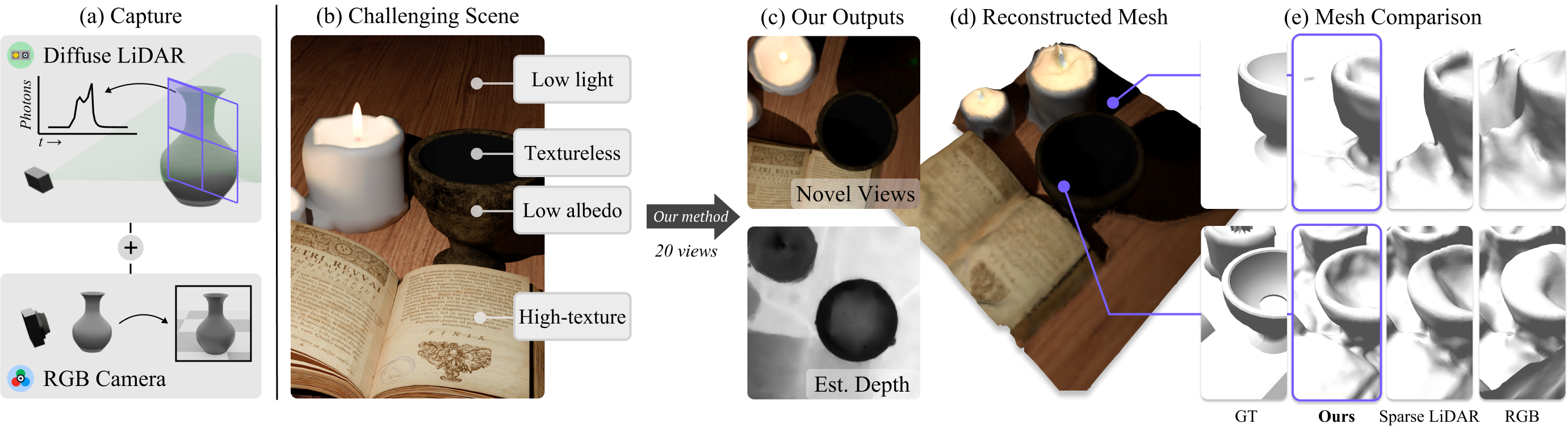} 
    \vspace{-4mm}
    \captionof{figure}{\textbf{Blurred LiDAR for Sharper 3D}: We propose leveraging (a) the complementary strengths of RGB with a diffuse (blurred) LiDAR for robust handheld scanning (b) on limited views in challenging low-texture, low-light, low-albedo scenes. Our method dynamically balances RGB and diffuse LiDAR sensor inputs to estimate (c) precise color, depth, and normals, from which we can (d) reconstruct accurate color-3D meshes. Our approach leveraging blurred LiDAR, counterintuitively, can (e) improve 3D reconstruction over conventional LiDAR, which uses sparse spot illumination. }
\end{center}%
}]

\maketitle

\begin{abstract}

\footnotetext[1]{\hspace{-1.5em} \textbf{Project Page:} \href{https://nikhilbehari.github.io/bls3d-web/}{nikhilbehari.github.io/bls3d-web/}}

\vspace{-1em}
\noindent 3D surface reconstruction is essential across applications of virtual reality, robotics, and mobile scanning. However, RGB-based reconstruction often fails in low-texture, low-light, and low-albedo scenes. Handheld LiDARs, now common on mobile devices, aim to address these challenges by capturing depth information from time-of-flight measurements of a coarse grid of projected dots. Yet, these sparse LiDARs struggle with scene coverage on limited input views, leaving large gaps in depth information. In this work, we propose using an alternative class of ``blurred" LiDAR that emits a \textit{diffuse flash}, greatly improving scene coverage but introducing spatial ambiguity from mixed time-of-flight measurements across a wide field of view. To handle these ambiguities, we propose leveraging the complementary strengths of diffuse LiDAR with RGB. We introduce a Gaussian surfel-based rendering framework with a scene-adaptive loss function that dynamically balances RGB and diffuse LiDAR signals. We demonstrate that, surprisingly, diffuse LiDAR can outperform traditional sparse LiDAR, enabling robust 3D scanning with accurate color and geometry estimation in challenging environments.

\vspace{-1em}

\end{abstract}    
\section{Introduction}
\label{sec:intro}

3D reconstruction has become increasingly important in applications such as virtual reality, mobile scanning, and robotics. In each such field, it is desirable to have compact, low-bandwidth, and low-cost capture hardware. Yet, simultaneously, 3D reconstruction must be robust to many challenging conditions, such as low-texture and low-albedo objects, and low-lighting scenes. Such qualities are common in real-world settings--for example, when a robot navigates an indoor space with textureless walls or when the Mars Helicopter scans dark, featureless sand dunes \cite{lorenz2023sounds}. Thus, an ideal setup combines lightweight, compact hardware with accurate, robust reconstruction capabilities. 

While recent work has extensively explored 3D reconstruction from RGB, these methods struggle to achieve robust performance in these challenging low texture, albedo, and lighting settings. For instance, recent work in Neural Radiance Fields (NeRF) ~\cite{mildenhall2021nerf} has enabled high-fidelity novel view synthesis in ideal settings; extensions have also demonstrated accurate reconstruction using signed distance fields~\cite{yariv2021volume} and Gaussian Surfels~\cite{dai2024high} for precise depth, normal and mesh estimation. However, NeRFs and Gaussian Splatting~\cite{kerbl20233d} primarily rely on multi-view appearance variations (e.g., texture) from RGB images; however, these images lack depth information in scenes with low-texture, low-lighting, or low-albedo. In this paper, we propose fusing RGB with an unconventional but widely-available sensor modality--diffuse LiDAR--for robust handheld 3D scanning in these challenging scene conditions.  



To improve reconstruction in these challenging settings, LiDAR is often used with RGB to enhance depth estimation in NeRF and other 3D reconstruction techniques. In handheld scanning, this is achieved by projecting a coarse grid of points into the scene and estimating precise depth at these locations. These depth values can directly supervise reconstruction models \cite{deng2022depth}. As these \textit{sparse} LiDARs are coupled with active illumination, they provide high signal-to-noise ratio (SNR) even in low-SNR conditions, like low lighting, making them effective for compact scanning setups (e.g. mobile phones \cite{rangwala2020,visionpro2024,sonyphone2021}). However, sparse LiDAR has a key limitation: each measurement captures only a coarse grid of individual depth points, resulting in poor scene coverage and therefore requiring many captures for adequate depth understanding. This trade-off between depth accuracy and sparse coverage can restrict its utility in settings where extensive multi-view captures are impractical.

In this work, we explore the benefits of an alternative diffuse LiDAR that, when paired with RGB, can surprisingly \textit{enhance} 3D reconstruction over sparse LiDAR in challenging conditions such as low-texture, low-light, and low-albedo. Unlike sparse LiDARs, which project individual dots to measure point depth, these LiDARs emit a \textit{diffuse} flash; each diffuse LiDAR pixel then captures a wide field-of-view, resulting in spatially blurred measurements. However, this lower depth precision comes with the benefit of much higher scene coverage. Our key insight is that, despite their spatially blurred measurements, these diffuse LiDARs can be leveraged in an analysis-by-synthesis framework to recover depth, enabling significantly improved scene reconstruction with limited views. A novel supervision strategy is needed, however, to integrate these diffuse LiDARs with RGB for reconstruction; unlike sparse LiDAR, direct point-wise depth supervision is infeasible from spatially blurred depth signals. Thus, we propose a scene-adaptive loss that dynamically balances RGB and diffuse LiDAR signals, prioritizing LiDAR in regions where RGB offers fewer multi-view cues.

\noindent \textbf{Our contributions in this work are as follows: }

\begin{itemize}
    \item  We propose leveraging diffuse LiDAR sensors with RGB for robust handheld 3D scanning in challenging low-texture, low-light, low-albedo scenarios. 
    \item We demonstrate that 1) diffuse LiDAR improves spatial coverage over conventional sparse lidar at the cost of spatial ambiguity, and 2) this spatial ambiguity can be resolved with RGB information (\cref{sec:formation}). 
    \item We propose a Gaussian-surfel 3D reconstruction technique and a scene-adaptive loss for balancing the complementary strengths of RGB and diffuse LiDAR (\cref{sec:method}).  
    \item We demonstrate the benefits of diffuse LiDAR over conventional sparse LiDAR through recoverability analysis (\cref{fig:analysis}), quantitative empirical evaluations (\cref{table:baseline-results}, \cref{table:ablation_lidaronly}, \cref{fig:lowlighting}), and qualitative real world experiments (\cref{fig:comparison_real}). 
\end{itemize}

\vspace{-4mm}

\paragraph{Scope of this Work.} We focus on commercial-grade LiDARs used in handheld settings, which have low spatial and temporal resolution. We consider 3D scanning of static, non-specular objects; future works could explore robustness to different materials and dynamic scenes. We also assume pose estimated from RGB -- future work could explore joint pose estimation with LiDAR.

\vspace{-2mm}

\section{Related Work}
\label{sec:relatedwork}



\begin{figure*}[!ht]
\centering
  \includegraphics[width=0.95\textwidth]{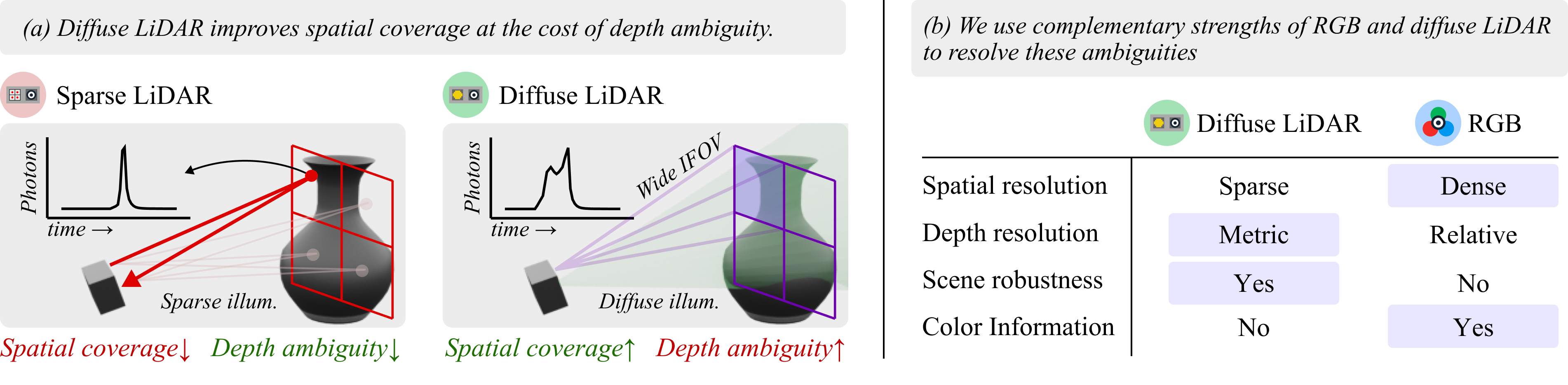} 
  \vspace{-4mm}
\caption{\textbf{Diffuse LiDAR + RGB for surface reconstruction}. (a) Sparse (conventional) LiDAR vs Diffuse LiDAR. Sparse LiDAR projects a grid of points which enable precise timing returns corresponding to individual depths; alternatively, diffuse LiDAR projects a diffuse flash illumination and measures the returns over a wide per-pixel instantaneous field-of-view (IFOV), increasing spatial coverage but also ambiguity in inferred depth. (b) Diffuse LiDAR and RGB have complementary strengths; RGB provides dense spatial and color information, while diffuse LiDAR provides coarse, metric depth even under challenging scenarios. } 
\label{fig:strengths_figure}
\vspace{-4mm}
\end{figure*}

\subsection{3D Neural Reconstruction}
NeRF has emerged as an effective approach for learning volumetric scene representations from 2D images \cite{mildenhall2021nerf}. Building on NeRF, methods for learning implicit surfaces enable surface geometry and normals to be learned  \cite{wang2021neus}. Finally, 3D Gaussian Splatting \cite{kerbl20233d} has been proposed for real-time rendering and, built on top of it,  2D ``Surfels" have enabled estimation of high-fidelity 3D surfaces \cite{dai2024high}.

Neural reconstruction has been widely applied beyond RGB cameras, e.g. sonar \cite{qadri2024aoneus}, CT \cite{gilo2024epinafenhancingneuralattenuation}, radar \cite{borts2024radar}, and LiDAR \cite{zhang2024nerf, wu2024dynamic, huang2023neural}. However methods that leverage time-of-flight sensors \cite{malik2024transient, luo2024transientangelo, malik2024flying} primarily focus on high-end, scanning laser setups, rather than emerging low-cost, small-form-factor, single-photon sensors. In addition, while fusing RGB with depth when training NeRF has been proposed \cite{deng2022depth,song2023darf}, these methods rely on RGB-based cues from COLMAP \cite{schonberger2016structure} or monocular depth estimation, which are limited in the low-texture, albedo, and illumination scenarios that we focus on.


\subsection{Time-of-Flight Imaging}
LiDARs emit short pulses of light and measure the \textit{time of flight} (ToF) of incident photons. While traditional LiDARs are used to measure only point depth, single-photon avalanche diodes (SPADs) are an increasingly common LiDAR sensor that record \textit{transient} images of photon intensity over time \cite{o2017reconstructing}. In addition to depth estimation \cite{jayasuriya2015depth, jungerman20223d}, SPADs have been used for a myriad of applications, such as measuring fluorescence lifetimes \cite{lee2023caspi}, non-line-of-sight (NLOS) imaging \cite{raskar2020seeing,kirmani2009looking,heide2019non}, and seeing through scattering media \cite{faccio2020non,xin2019theory,velten2012recovering}. Recently, NeRF-based methods have been developed for 3D reconstruction  \cite{klinghoffer2024platonerf, attal2021torf, malik2024transient, okunev2025flowed, jungerman2024radiance} and NLOS imaging \cite{shen2021non, mu2022physics, fujimura2023nlos} from SPADs. Whereas most methods rely on high-cost or laboratory-grade hardware setups, the focus of our work is to demonstrate the utility of low-cost SPADs that use diffuse, rather than point-based, illumination. While recent work has shown promising results with these sensors \cite{sifferman2023unlocking, mu2024towards}, none have explored the complementary strengths of low-cost SPADs with RGB sensors. While SPADs and RGB have been used together for single-view depth estimation \cite{li2022deltar, meuleman2022floatingfusion,sun2023consistent}, we focus on multi-view 3D reconstruction across challenging scenes.

\section{Leveraging Diffuse LiDAR and RGB}

\label{sec:formation}

LiDAR is widely used with RGB for surface reconstruction in robotics and handheld scanning. Although RGB provides dense spatial and color information, it struggles with low-light, low-texture, and low-albedo scenes. In these challenging settings, LiDAR can improve reconstruction by providing metric depth at discrete sparse locations; however, with limited input views, conventional LiDAR may lack the scene coverage needed for accurate reconstruction. In this section, we discuss depth supervision using an alternative class of LiDAR, diffuse LiDARs. These LiDARs greatly improve spatial coverage over conventional LiDARs, but also introduce spatial blur into the imaging process. We analytically show how diffuse LiDARs, counterintuitively, can improve recoverability over sparse LiDAR despite their spatial blur. We then offer insights for leveraging the combined strengths of diffuse LiDAR and RGB, particularly for challenging settings where RGB alone may not be sufficient for accurate 3D reconstruction (\cref{fig:strengths_figure}).

\subsection{LiDAR for Improved Handheld 3D Scanning}
\label{sec:sparse_lidar}

LiDARs are widely used to improve 3D reconstruction by providing point depth supervision. They operate with a co-located illumination source and a detector, as shown in \cref{fig:strengths_figure}(a). The laser emits a pulse of light towards a scene point $\mathbf{x}$. The time it takes for that pulse of light to travel along the path from the $\text{laser} \rightarrow \mathbf{x} \rightarrow \text{detector}$ is $t_\mathbf{x} = 2\left\|\mathbf{x} \right \| / c$, where $c$ is the speed of light. The ToF measurement can be expressed as

\vspace{-1mm}
\begin{equation}
    i_{\mathbf{x}}(t) = \delta(t - t_\mathbf{x}),
    \label{eq:tof}
\end{equation}

\noindent where $\delta(\cdot)$ is the delta function modeling the time delay based on the camera distance to $\mathbf{x}$. From a ToF measurement, we can estimate the $t_\mathbf{x}$, from which we can directly estimate the depth of a scene point.



Although conventional LiDARs provide benefits for 3D scanning, they have several limitations, particularly when RGB is also insufficient for reconstruction. First, each pixel images exactly one scene point, which means that a large number of pixels are required to obtain sufficient spatial coverage of the scene. Second, most solid-state LiDAR systems are based on avalanche diodes, which have substantial power requirements. Third, LiDARs based on time-resolved imaging output high-dimensional data structures, which incurs a bandwidth cost. Taken together, these limitations impose a tradeoff for lightweight mobile applications: LiDARs require a large number of pixels to obtain sufficient spatial coverage, but increasing the number of pixels induces a power and bandwidth penalty \cite{zhang2022first}.


\subsection{Diffuse LiDAR Enables Better Coverage}
\label{sec:diffuse_lidar}



We aim to leverage the 3D capabilities of LiDARs, without making large concessions on spatial information, bandwidth, or power. Our key insight is that we can utilize \textit{diffuse LiDARs}, which encode more information into a single pixel measurement by (1) increasing the pixel's instantaneous field of view (IFOV) and (2) using a diffused light source, as shown in \cref{fig:strengths_figure}(b). In this case, the pixel transient measurement can be expressed as

\begin{equation}
    i(t) = \int_{\mathbf{x} \in \Omega} i_{\mathbf{x}}(t) d\mathbf{x},
\end{equation}

\noindent where $\Omega$ is the set of 3D surface points $\mathbf{x}$ in the pixel's IFOV. 

The key benefit of this capture setup is that each sensor pixel can capture information from multiple scene points \emph{simultaneously}. Therefore, fewer pixels are needed to capture information from all scene points. The challenge now lies in separating light coming from different scene points. In a conventional camera, large IFOV typically results in pixel blur that cannot be recovered. However, a LiDAR can measure the ToF of incident light, providing a key source of information in disambiguating light contributions from different scene points. A diffuse LiDAR's ability to decompose light contributions from different scene points is partially determined by the number of views \cite{somasundaram2023role}, which motivates an analysis on when diffuse LiDAR is beneficial. 

\begin{figure}[!tb]
\centering
\includegraphics[width=\linewidth]{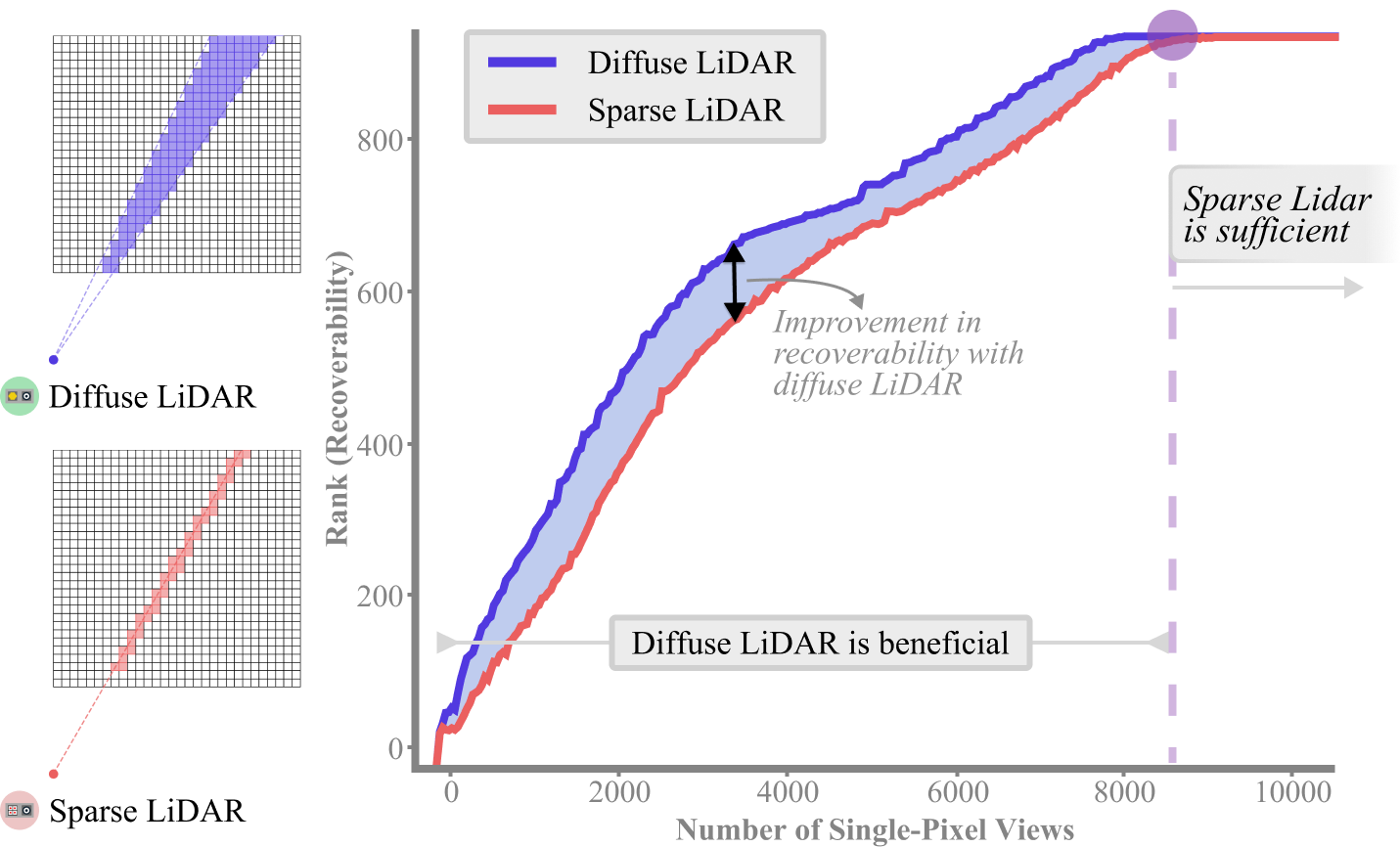}
\vspace{-4mm}
\caption{\textbf{Improved recoverability with diffuse LiDAR when input views are limited}. Full rank is 900 in our analysis simulation. Diffuse LiDAR has greater voxel coverage than conventional sparse LiDAR; this greater coverage can improve rank, and thereby recoverability, when using a limited number of input views. We consider this specific limited-view domain in this work. As the number of input views increases, sparse LiDAR can eventually provide sufficient coverage for scene recoverability.}
\label{fig:analysis}
\vspace{-6mm}
\end{figure}

\label{sec:bluranalysis}
\vspace{1mm}
\noindent
\textbf{Analysis of Scene Recoverability.} To quantify the benefits of diffuse LiDAR, we perform an analysis to compare diffuse and sparse LiDARs. For computational and conceptual simplicity, we consider a 2D scene without loss of generality. Our analysis consists of an approximate linear model (details in Supplementary) to map voxelized scene geometry $\mathbf{x}$ to LiDAR measurements $\mathbf{y}$ as described by: $\mathbf{y}=\mathbf{A}\mathbf{x}$. The size and structure of $\mathbf{A}$ is a function of the number of views and the IFOV of the LiDAR. As a result, by computing matrix properties (i.e. rank) of $\mathbf{A}$, we can quantify the recoverability of a scene under different LiDAR capture configurations. In \cref{fig:analysis}, we plot the rank of $\mathbf{A}$ for a diffuse LiDAR (high IFOV) and a sparse LiDAR (low IFOV) as we increase the number of views of the scene.  We observe that the rank increases with number of views for both configurations, but the diffuse LiDAR offers an improvement in recoverability due to its spatial coverage of the scene. In spite of the improvement over sparse LiDAR, diffuse LiDAR alone is still insufficient for scene recovery when using fewer views and low spatial resolution, which motivates our RGB-SPAD fusion in \cref{sec:method}.

\subsection{Balancing Diffuse LiDAR \& RGB}
As discussed above, diffuse LiDAR has potential to improve depth recoverability over sparse LiDAR in limited-view scenarios; however, it also introduces ambiguity due to spatially mixed measurements encoded in time. To address these ambiguities, we leverage the complementary strengths of RGB sensors, which provide dense spatial and color data but struggle in low-light, low-texture conditions (see \cref{fig:strengths_figure}b). Our method aims to fuse both sensors to handle scenes ranging from ideal to challenging. Yet, unlike sparse LiDAR, diffuse LiDAR captures depth across a wide field of view, making direct point-wise depth supervision used in sparse LiDAR supervision infeasible. Thus, a new rendering approach is required to effectively combine these complementary signals for accurate 3D reconstruction.

\begin{figure*}[!ht]
  \centering
  \includegraphics[width=1.0\textwidth]{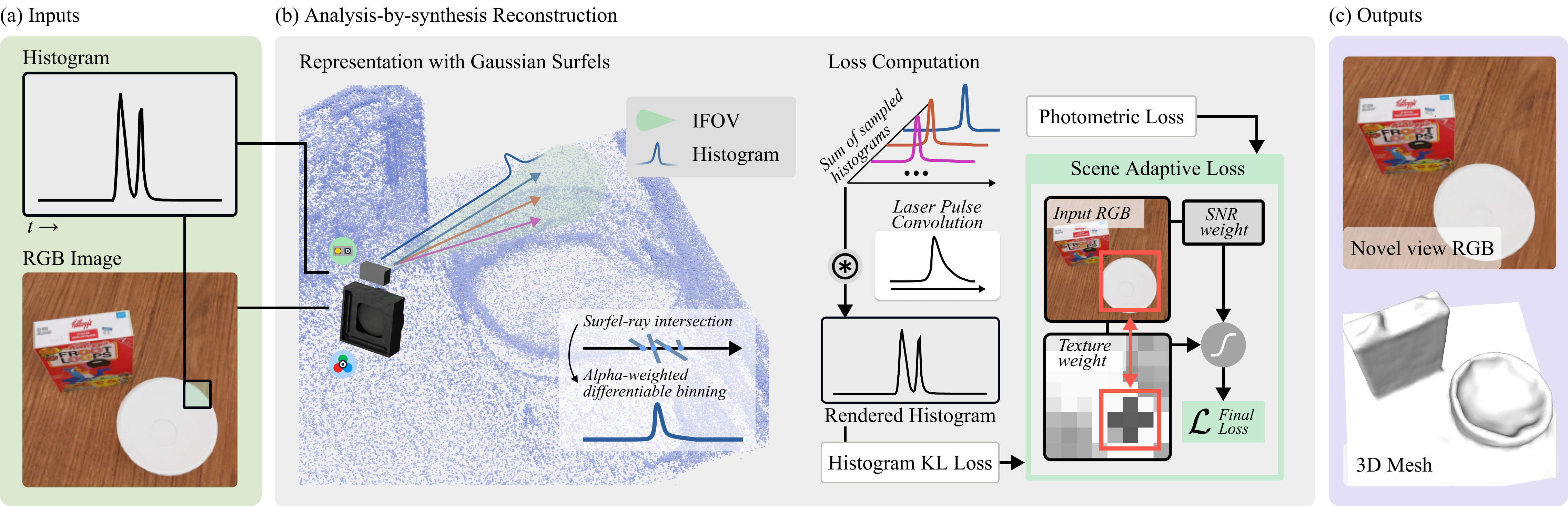}
  \vspace{-5mm}
   \caption{\textbf{Inverse rendering with RGB and diffuse LiDAR}. We consider a compact hardware setting with co-located RGB camera and diffuse LiDAR. At each view, we capture an (a) RGB image and coarse ($8 \times 8$) histograms, for which each pixel contains mixed signal from a wide IFOV $\omega$. We perform (b) analysis-by-synthesis reconstruction using Gaussian surfels, sampling rays within each pixel IFOV and rendering transients with alpha-weighted differentiable binning. Loss signal from RGB and transient inputs are balanced dynamically with a scene-adaptive loss producing (c) high-fidelity RGB and depth/normals for accurate mesh reconstruction. }
  \label{fig:methods}
  \vspace{-4mm}
\end{figure*}

\section{Reconstruction with Diffuse LiDAR \& RGB}

\label{sec:method}

In this section, we describe an inverse rendering procedure that recovers surface geometry and texture from multi-view RGB and diffuse LiDAR measurements. The RGB and LiDAR sensors are rigidly mounted together with known relative pose between the two sensors. We use COLMAP \cite{schonberger2016structure} to obtain the camera intrinsics and per-frame extrinsics. 

Our analysis-by-synthesis reconstruction method is based on a Gaussian surfel scene representation \cite{dai2024high} and a differentiable RGB-transient rendering algorithm. By using a scene-adaptive loss function, we can dynamically leverage RGB signal in high-texture, high SNR settings, and gradually de-emphasize RGB and prioritize diffuse LiDAR signals in low-texture and low-SNR regions. 



\subsection{Surfel-Based RGB-SPAD Rendering}

\textbf{Gaussian Surfels.} We represent the 3D scene as a composition of Gaussian surfel primitives \cite{dai2024high}. Surfels are similar to the 3D Gaussian representation used in GS \cite{kerbl20233d}, with the key difference being that the 3D Gaussian is flattened into a 2D Gaussian. These 2D Gaussians can be oriented to align with the surface of a 3D object. 

The shape and location of the Gaussian can be parametrized by a covariance matrix $\mathbf{\Sigma} \in \mathbb{R}^{3 \times 3}$ and mean vector $\mathbf{x}_i \in \mathbb{R}^3$ respectively. The resulting 3D Gaussian can be expressed as  
\vspace{-1mm}
\begin{equation}
G(\mathbf{x}; \mathbf{x}_i, \mathbf{\Sigma}_i) = e^{ -\frac{1}{2} (\mathbf{x} - \mathbf{x}_i)^\top \mathbf{\Sigma}_i^{-1} (\mathbf{x} - \mathbf{x}_i) }.
\end{equation}
\vspace{-1mm}
\noindent Covariance matrices, by definition, are positive semi-definite and can be interpreted as describing the orientation of an ellipsoid. To enforce both of these properties, the covariance matrix can be decomposed as 
\vspace{-1mm}
\begin{equation}
    \mathbf{\Sigma} = (\mathbf{R}\mathbf{S})(\mathbf{R}\mathbf{S})^\top,
\end{equation}
\vspace{-1mm}
\noindent where $\mathbf{R} \in \mathbb{R}^{3 \times 3}$ is a rotation matrix and $\mathbf{S} \in \mathbb{R}^{3 \times 3}$ is a diagonal scaling matrix describing the scaling of the principal axes. $\mathbf{R}$ can be analytically expressed in terms of a quaternion $\mathbf{r} \in \mathbb{R}^4$ and the scaling matrix can be expressed as $\mathbf{S}=\operatorname{diag}(\mathbf{s})$, where $\mathbf{s} \in \mathbb{R}^3$ is the scaling vector. To flatten the 3D Gaussian to a 2D surfel, we set $\mathbf{s} = [s_1, s_2, 0]^\top$.

The resulting parametrization of the Gaussian kernels $\{\mathbf{x}_i, \mathbf{r}_i, \mathbf{s}_i, o_i, \mathcal{C}_i\}$ has five degrees of freedom: the mean $\mathbf{x}_i \in \mathbb{R}^3$ is the 3D location, $\mathbf{r}_i \in \mathbb{R}^4$ is the rotation expressed as a quaternion, $\mathbf{s} \in \mathbb{R}^3$ is scaling, $o_i \in \mathbb{R}$ is the opacity, and $\mathcal{C}_i \in \mathbb{R}^k$ are spherical harmonic coefficients.



\vspace{2mm}
\noindent
\textbf{Surfel Rasterization.} We follow prior works in volumetric rendering to rasterize surfel primitives to the image plane \cite{dai2024high, kerbl20233d, zwicker2001ewa}. The projection of a 3D Gaussian onto a 2D image plane can be approximated as a 2D Gaussian with mean $\mathbf{u}_i \in \mathbb{R}^2$ and covariance $\mathbf{\Sigma}_i' \in \mathbb{R}^{2 \times 2}$
\vspace{-1mm}
\begin{equation}
G'(\mathbf{u}; \mathbf{u}_i, \mathbf{\Sigma}'_i) = e^ {-\frac{1}{2} (\mathbf{u} - \mathbf{u}_i)^\top \mathbf{\Sigma}_i'^{-1} (\mathbf{u} - \mathbf{u}_i) }.
\end{equation}
\vspace{-1mm}
\noindent Given a view transformation $\mathbf{W} \in \mathbb{R}^{3\times 3}$ and the Jacobian of the affine approximation of the projective transformation $\mathbf{J} \in \mathbb{R}^{3\times 3}$, the covariance in image space is approximately 

\vspace{-1mm}
\begin{equation}
\mathbf{\Sigma}'_i = \left( \mathbf{J} \mathbf{W} \mathbf{\Sigma}_i \mathbf{W}^\top \mathbf{J}^\top \right) [: 2, : 2].
\end{equation}
\vspace{-3mm}

\noindent
\textbf{RGB Rendering.} The color $\tilde{\mathbf{C}}$ at a pixel $\mathbf{u}$ can be computed by sampling all $n$ surfels along the camera ray corresponding to the pixel. These samples can then be integrated to render a color using the alpha compositing equation

\vspace{-1mm}
\begin{equation}
\tilde{\mathbf{C}} = \sum_{i=0}^{n} T_i \alpha_i \mathbf{c}_i, \ \ T_i = \prod_{j=0}^{i-1} (1 - \alpha_j),  
\end{equation}
\vspace{-1mm}

\noindent where $\alpha_i = G'(\mathbf{u}; \mathbf{u}_i, \Sigma'_i) o_i$ is the alpha-blending weight and $\mathbf{c}_i$ is the color computed using $\mathcal{C}_i$ \cite{kerbl20233d}.

\vspace{2mm}
\noindent
\textbf{Surface Rendering.} The depth $\tilde{D}$ at a pixel can be estimated by computing the expected distance along the ray 

\vspace{-1mm}
\begin{equation}
    \tilde{D} = \frac{1}{1 - T_{n+1}} \sum_{i=0}^{n} T_i \alpha_i d_i(\mathbf{u}),
    \label{eq:render_depth}
\end{equation}
\vspace{-1mm}

\noindent where $d_i(\mathbf{u})$ is the distance of the $i$th Gaussian from the camera along pixel direction $u$. A key benefit of the surfel representation is that depth can be estimated precisely for any 2D surface in 3D space \cite{dai2024high}. The resulting intersection between the pixel ray and $i$th surfel can be estimated as  

\vspace{-2mm}
\begin{equation}
    d_i(\mathbf{u}) = d_i(\mathbf{u}_i) + (\mathbf{W} \mathbf{R}_i)[2, :] \mathbf{J}_{pr}^{-1} (\mathbf{u} - \mathbf{u}_i),
    \label{eq:surfel_depth}
\end{equation}
\vspace{-2mm}

\noindent where $\mathbf{J}_{pr}^{-1}$ is the Jacobian of inverse mapping, and $(\mathbf{W} \mathbf{R}_i)$ transforms the rotation matrix to the camera space. The surface normal $\tilde{\mathbf{N}}$ can be computed by replacing $d_i(\mathbf{u})$ in \cref{eq:render_depth} with the $z$ direction of the rotation matrix $\mathbf{R}_i[:, 2]$.

\begin{table*}[!ht]
\centering
\caption{\textbf{Quantitative evaluation of low-texture scene reconstruction on rendered scenes.} We compare our method to: 1) Gaussian surfels (RGB Only), 2) Surfels with RGB and monocular depth, and 3) Surfels with RGB and sparse LiDAR. We evaluate for 10 input images on four scenes, each with four texture variations. We obtain consistent improvement in depth (D.MAE $\downarrow$) and normal (N.MAE $\downarrow$) estimation across texture variations, improving robustness over conventional LiDAR in low-texture, limited-view settings.}
\vspace{-3mm}
\label{table:baseline-results}
\scriptsize
\resizebox{\textwidth}{!}{
\begin{tabular}{l | cccc| cccc| cccc| cccc}
\toprule
\multirow{2}{*}{\textbf{Method}} & \multicolumn{4}{c|}{\textbf{Blender}} & \multicolumn{4}{c|}{\textbf{Chair}} & \multicolumn{4}{c|}{\textbf{Hotdog}} & \multicolumn{4}{c}{\textbf{Lego}} \\
& \textbf{PSNR ↑} & \textbf{SSIM ↑} & \textbf{D.MAE ↓} & \textbf{N.MAE ↓} & \textbf{PSNR ↑} & \textbf{SSIM ↑} & \textbf{D.MAE ↓} & \textbf{N.MAE ↓} & \textbf{PSNR ↑} & \textbf{SSIM ↑} & \textbf{D.MAE ↓} & \textbf{N.MAE ↓} & \textbf{PSNR ↑} & \textbf{SSIM ↑} & \textbf{D.MAE ↓} & \textbf{N.MAE ↓} \\
\toprule
\rowcolor{lightgray!50} \multicolumn{17}{l}{\textbf{(a) Full Texture Datasets}: Scenes with full texture on object and ground plane.} \\
\midrule
Surfels RGB Only & 28.92 & 0.848 & 0.081 & 34.27 & 28.68 & 0.852 & 0.062 & 28.95 & 30.85 & 0.859 & 0.068 & 36.12 & 27.67 & 0.794 & 0.074 & 38.40 \\
Surfels RGB w/ Mono. Depth & 27.77 & 0.876 & \secon 0.033 & \first 17.50 & 27.94 & 0.866 & \secon 0.028 & \first 16.06 & 28.97 & 0.886 & \secon 0.031 & \first 20.75 & 27.42 & 0.818 & \secon 0.032 & \first 21.64 \\
Surfels RGB w/ Sparse LiDAR & 24.68 & 0.839 & \third 0.058 & \third 24.74 & 26.06 & 0.840 & \third 0.057 & \third 23.44 & 27.85 & 0.856 & \third 0.054 & \third 29.49 & 26.12 & 0.801 & 0.058 & 30.63 \\
\textbf{Ours (RGB + Diffuse LiDAR)} & 30.67 & 0.881 & \first 0.025 & \secon 19.17 & 30.25 & 0.891 &  \first 0.017 & \secon 18.05 & 30.14 & 0.894 & \first 0.016 & \secon 24.62 & 28.39 & 0.845 & \first 0.025 & \secon 25.27 \\

\toprule
\rowcolor{lightgray!50} \multicolumn{17}{l}{\textbf{(b) Textured Object Datasets}: Scenes with textured objects on completely textureless ground planes.} \\
\midrule
Surfels RGB Only             & 25.44 & 0.944 & 0.226 & 50.88 & 22.89 & 0.928 & 0.474 & 59.75 & 37.02 & 0.962 & 0.104 & 41.49 & 26.06 & 0.878 & 0.176 & 54.25 \\
Surfels RGB w/ Mono. Depth       & 27.66 & 0.952 & \third 0.129 & \secon 28.91 & 21.99 & 0.932 & \third 0.452 & \third 55.76 & 36.04 & 0.961 & \secon 0.065 & \secon 26.33 & 27.65 & 0.888 & \secon 0.052 & \secon23.67 \\
Surfels RGB w/ Sparse LiDAR      & 26.96 & 0.937 & \secon 0.105 & \third 36.91 & 25.64 & 0.936 & \secon 0.115 & \secon38.68 & 27.03 & 0.932 & \third 0.080 & \third 34.98 & 25.77 & 0.879 & \third 0.093 & \third 39.99 \\
\textbf{Ours (RGB + Diffuse LiDAR)}  & 25.08 & 0.945 & \first 0.033 & \first 10.47 & 32.18 & 0.970 & \first 0.030 & \first 7.33 & 35.62 & 0.963 & \first 0.024 & \first 16.34 & 30.00 & 0.924 & \first 0.024 & \first 16.94 \\
\toprule
\rowcolor{lightgray!50} \multicolumn{17}{l}{\textbf{(c) Textured Plane Datasets}: Scenes with completely textureless objects on textured ground planes. } \\
\midrule
Surfels RGB Only            & 26.34 & 0.848 & 0.090 & 37.53 & 23.69 & 0.841 & 0.089 & 36.90 & 29.30 & 0.879 & 0.101 & 44.77 & 25.86 & 0.858 & 0.091 & 42.49 \\
Surfels RGB w/ Mono. Depth      & 26.34 & 0.883 & \first 0.036 & \first 19.27 & 23.76 & 0.867 & \secon 0.049 & \first 20.61 & 30.41 & 0.921 & \third 0.050 & \secon 27.08 & 24.95 & 0.889 & \secon 0.055 & \secon 28.24 \\
Surfels RGB w/ Sparse LiDAR     & 25.36 & 0.867 & \third 0.057 & \third 23.95 & 24.92 & 0.842 & \third 0.063 & \third 31.37 & 28.33 & 0.894 & \secon 0.045 & \third 29.28 & 24.69 & 0.861 & \third 0.067 & \third 31.06 \\
\textbf{Ours (RGB + Diffuse LiDAR)} & 23.72 & 0.820 & \secon 0.045 & \secon 21.13 & 25.70 & 0.841 & \first 0.037 & \secon 22.80 & 28.97 & 0.858 &\first  0.034 & \first 22.32 & 26.60 & 0.850 & \first 0.046 & \first 25.69 \\
\toprule
\rowcolor{lightgray!50} \multicolumn{17}{l}{\textbf{(d) No Texture Datasets}: Completely textureless scenes. } \\
\midrule
Surfels RGB Only            & -- & -- & 0.885 & \secon 36.74 & -- & -- & 0.904 & \third 42.14 & -- & -- & 0.850 & 63.31 & -- & -- & 0.898 & \third 51.27 \\
Surfels RGB w/ Mono. Depth      & -- & -- & \third 0.716 & 67.01 & -- & -- & \third 0.793 & 60.23 & -- & -- & \third 0.911 & \third 46.24 & -- & -- & \third 0.835 & 63.31 \\
Surfels RGB w/ Sparse LiDAR     & -- & -- & \secon 0.125 & \third 40.86 & -- & -- & \secon 0.107 & \secon 37.29 & -- & -- & \secon 0.111 & \secon 43.54 & -- & -- & \secon 0.106 & \secon 41.63 \\
\textbf{Ours (RGB + Diffuse LiDAR)} & -- & -- & \first 0.045 & \first 15.68 & -- & -- & \first 0.045 & \first 13.55 & -- & -- & \first 0.041 & \first 16.35 & -- & -- & \first 0.042 & \first 14.54 \\
\toprule
\end{tabular}
}
\vspace{-6mm}
\end{table*}

\vspace{2mm}
\noindent
\textbf{Transient Rasterization.} We now synthesize ToF measurements from surfel primitives. A ToF measurement $i[x, y, t] \in \mathbb{R}^{N_x \times N_y \times N_t}$ is a 3D data structure, where $N_x$ and $N_y$ are the number of pixels in the $x$ and $y$ direction, and $N_t$ is the number of timing bins. Each pixel measurement is discretized into a histogram with $N_t$ bins, where the bin width is the timing resolution $\Delta t$ of the LiDAR. In our problem setting, we use larger pixels, which introduces spatial blur. As a result, each pixel measures radiance along a \emph{cone}, rather than along an individual ray. 

In order to render the histogram measurement $i[t]$ at a pixel $\mathbf{u}=[x, y]$, we first sample a set of rays $\mathcal{R}$ within the pixel cone, and compute the set of all surfels $\mathcal{S}_\mathbf{r}$ intersecting each ray $\mathbf{r} \in \mathcal{R}$. For each surfel $s \in \mathcal{S}_\mathbf{r}$, we assign its radiance contribution to a temporal histogram bin index $\beta_s$  

\vspace{-1mm}
\begin{equation}
\beta_s = \left\lfloor \frac{2d_s}{\Delta t} \right\rfloor,
\end{equation}
\vspace{-1mm}

\noindent where $d_s$ is the depth to the surfel $s$ that can be computed using \cref{eq:surfel_depth}. Computation of the bin index can be interpreted as a temporal quantization of \cref{eq:tof}. In practice, we find that performing soft histogramming similar to prior work \cite{malik2024transient} improves gradient flow during rendering. The idea is to distribute the surfel contribution over bins $\beta_s$ and $\beta_s + 1$, with respective weights

\vspace{-2mm}
\begin{equation}
w_{\beta_{s,1}} = 1 - \left( \frac{d_s}{\Delta t} - \beta_s \right), w_{\beta_{s,2}} = 1 - w_{\beta_{s,1}}.
\end{equation}
\vspace{-2mm}

\noindent The final rendered histogram $i[t]$, for each bin $b$ at a pixel, is constructed by summing the opacity-weighted contributions across all surfels:

\vspace{-1mm}
\begin{equation}
i[t] = \sum_{\mathbf{r} \in \mathcal{R}} \sum_{s \in \mathcal{S}} o_s \cdot \left( w_{\beta_{s,1}} \delta[t- \beta_s] \right. \left. + w_{\beta_{s,2}} \delta[t - \beta_s - 1] \right).
\end{equation}
\vspace{-1mm}


\vspace{-4mm}
\subsection{Scene-Adaptive Loss Function}
The key benefit of sensor fusion in this work is that the shortcomings of one sensor can be overcome by relying on information from the other  (\cref{fig:strengths_figure}). In order to adaptively do so during optimization, we quantify the ``usefulness'' of information available in the RGB image. The loss function will use this quantification of usefulness to adaptively determine whether to rely on the RGB or LiDAR image more.

\vspace{2mm}
\noindent
\textbf{Quantifying Usefulness of RGB.} There are two key aspects of usefulness that we consider in the RGB images: texture and signal-to-noise ratio (SNR). Intuitively, RGB images contain more information when they contain textured scenes with sufficient SNR. We divide the RGB images into a set of patches $\mathcal{P}$ such that each patch $p$ corresponds to the IFOV of exactly one diffuse LiDAR pixel. We then compute per-patch values for SNR and texture. The SNR of an image patch $p$ can be computed as $w_{snr} = \mu_p /{\sigma^2_p}$, where $\mu_p$ and $\sigma^2_p$ are the mean and variance of the pixel intensities in the patch. The amount of texture within a patch can be quantified as the variance $w_{texture}=\sigma^2_p$. We use these weights as input to a sigmoid function 

\vspace{-1mm}
\begin{equation}
w_p(x, \epsilon, k) = \frac{1}{1 + e^{-k(x - \epsilon)}},
\end{equation}
\vspace{-1mm}

\noindent to determine the usefulness of RGB. We set input $x = w_{texture}$, translation parameter to be $\epsilon = aw_{snr}+b$, and $a$, $b$, and steepness parameter $k$ to be hyperparameters. This sigmoid weight helps us determine the weighting of the RGB patch, where high variance leads to higher weighting and low SNR leads to lower weighting. 

\begin{figure*}[ht!]
  \centering
  \includegraphics[width=\textwidth]{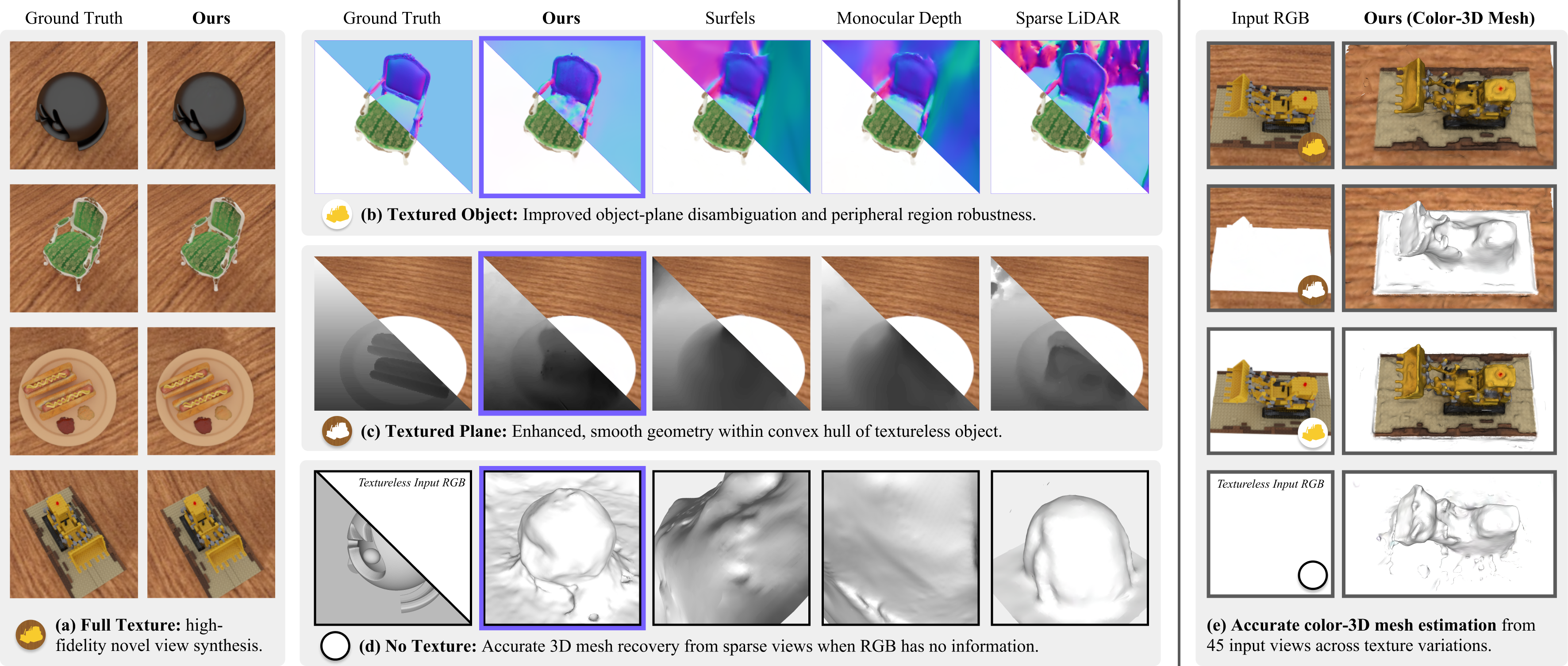}
  \vspace{-6mm}
  \caption{\textbf{Qualitative comparisons on rendered scenes with varying texture.} We enable (a) accurate RGB novel-view synthesis on full-texture scenes, where RGB may be prioritized by our adaptive loss. When (b) textured objects are on textureless planes, we enable greater object-plane separation and peripheral region robustness; on textureless objects, we (c) enhance geometry estimation within the object convex hull visible from RGB. We also (d) improve over sparse LiDAR in scenes without RGB signal, where our adaptive loss may weight diffuse LiDAR more heavily. We enable (e) accurate color-mesh estimation across a wide range of texture and object variations. }
  \vspace{-4mm}
  \label{fig:comparison_sim}
\end{figure*}

\vspace{2mm}
\noindent
\textbf{RGB Loss.} The RGB loss consists of two terms: data fidelity (L1) and perceptual similarity (SSIM)



\begin{equation}
\begin{split}
    \mathcal{L}_{\text{RGB}} 
    = (1 - \lambda_{\text{SSIM}}) \cdot \sum_{p \in \mathcal{P}}  w_p \| \tilde{\mathbf{C}}_p - \mathbf{C}_{p,gt} \|_1  \\
    + \lambda_{\text{SSIM}} \cdot (1 - \text{SSIM}(\tilde{\mathbf{C}}, \mathbf{C}_{gt})) \cdot \frac{1}{|\mathcal{P}|} \sum_{p \in \mathcal{P}} w_p.
\end{split}
\end{equation}

\vspace{-1mm}

\noindent $\mathbf{C}$ is the full image, $\mathbf{C}_p$ is the patch image, and
$\lambda_{\text{SSIM}}$ is a hyper-parameter weight. The L1 term is computed patch-wise, where each patch is weighted differently. The SSIM term is computed for the entire image and is weighted by the average weight across all image patches. 

\vspace{2mm}
\noindent
\textbf{Transient Loss.} The transient loss is the KL divergence between the normalized rendered transient and ground truth. 

\vspace{-3mm}
\begin{equation}
\mathcal{L}_{\text{transient}} = \sum_{p \in \mathcal{P}}(1-w_p) \cdot \text{KL}(i_p[t] \parallel i_{p, gt}[t]),
\end{equation}

\noindent where $i_p[t]$ is the transient corresponding to RGB patch $\mathcal{P}$.   



\noindent
\paragraph{Combined Loss Function.} The three terms in the final loss function are the RGB loss $\mathcal{L}_{rgb}$, the LiDAR loss $\mathcal{L}_{lidar}$, and a depth-normal consistency regularization $\mathcal{L}_{reg}$ \cite{dai2024high}. Putting these terms together results in the combined loss 

\vspace{-1mm}
\begin{equation}
    \mathcal{L} =\mathcal{L}_{rgb} + \mathcal{L}_{lidar} + \mathcal{L}_{reg}.
\end{equation}
\vspace{-1mm}

\vspace{-4mm}
\section{Experiments}
\label{sec:experiments}
\vspace{-1mm}

\textbf{Comparisons.}
We compare our method against three baseline techniques: 1) Gaussian surfels rendering \cite{dai2024high} using only RGB inputs; 2) Gaussian surfels with monocular depth priors \cite{yang2024depth}, which \cite{dai2024high} suggest can aid in surface reconstruction but that we expect to fail on low-texture scenes; 3) Gaussian surfels with sparse LiDAR depth loss. For the sparse LiDAR baseline, we apply the standard L1 depth loss commonly used in LiDAR-supervised NeRF methods (\( L_{\text{depth}}(\mathbf{r}) = \sum || D_{\text{est}}(\mathbf{r}) - D_{\text{true}}(\mathbf{r}) ||_1 \)).

\begin{figure}[t!]
\centering
\includegraphics[width=\linewidth]{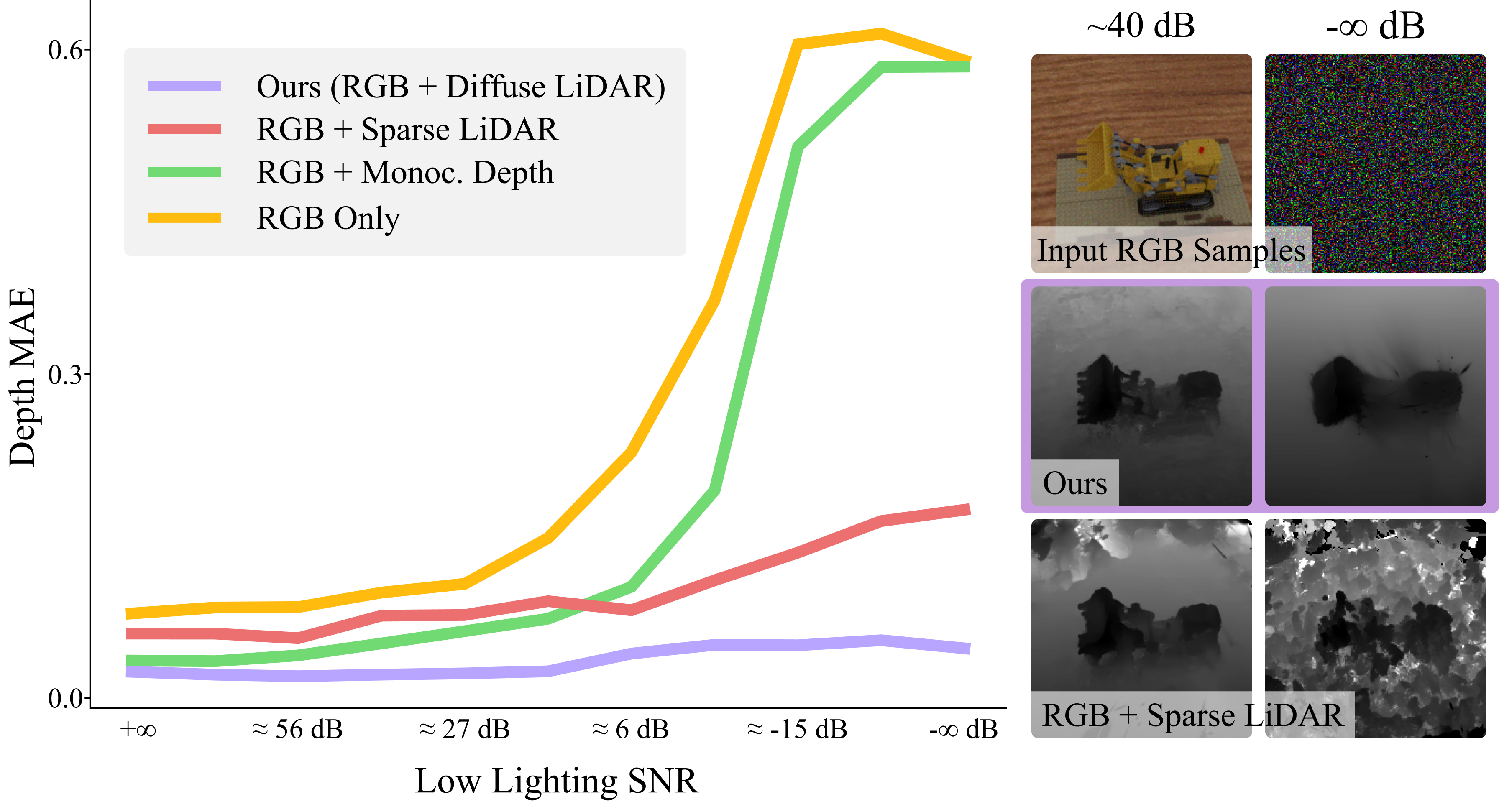}
\vspace{-6mm}
\caption{\textbf{Robust depth (MAE $\downarrow$) of our method in low lighting}. We simulate low lighting with added Gaussian noise; our \textit{scene-adaptive loss weighting} can be used to to rely on diffuse LiDAR inputs more heavily as RGB input SNR increases, enabling robust depth estimation across a wide scale of low lighting noise.}
\label{fig:lowlighting}
\vspace{-6mm}
\end{figure}

\subsection{Simulated Experiments}
\label{sec:sim-experiments}
\vspace{-2mm}

\textbf{Experimental Setup.} We simulate a diffuse LiDAR with an $8 \times 8$ pixel array, using approximate horizontal/vertical IFOV of $4.9^\circ$, maximum measurement distance 1.5 meters, and 40 picosecond bin resolution. We compare to spare LiDAR-supervised Surfels using 8x8 points located at the center of each diffuse LiDAR pixel zone. We render RGB and 8x8 diffuse LiDAR histograms for four scenes using the image formation model described in \cref{sec:method}: Blender Ball, Chair, Hotdog, Lego. To evaluate robustness to low and no-texture scenes, we consider four dataset variations: 1) Full Texture; 2) Textured Object; 3) Textured Plane; and 4) No Texture. Examples of these variations are shown for Lego in \cref{fig:comparison_sim}e. We consider a view-limited regime with 10 training and 10 test captures. 

\begin{figure}[!t]
\centering
\includegraphics[width=\linewidth]{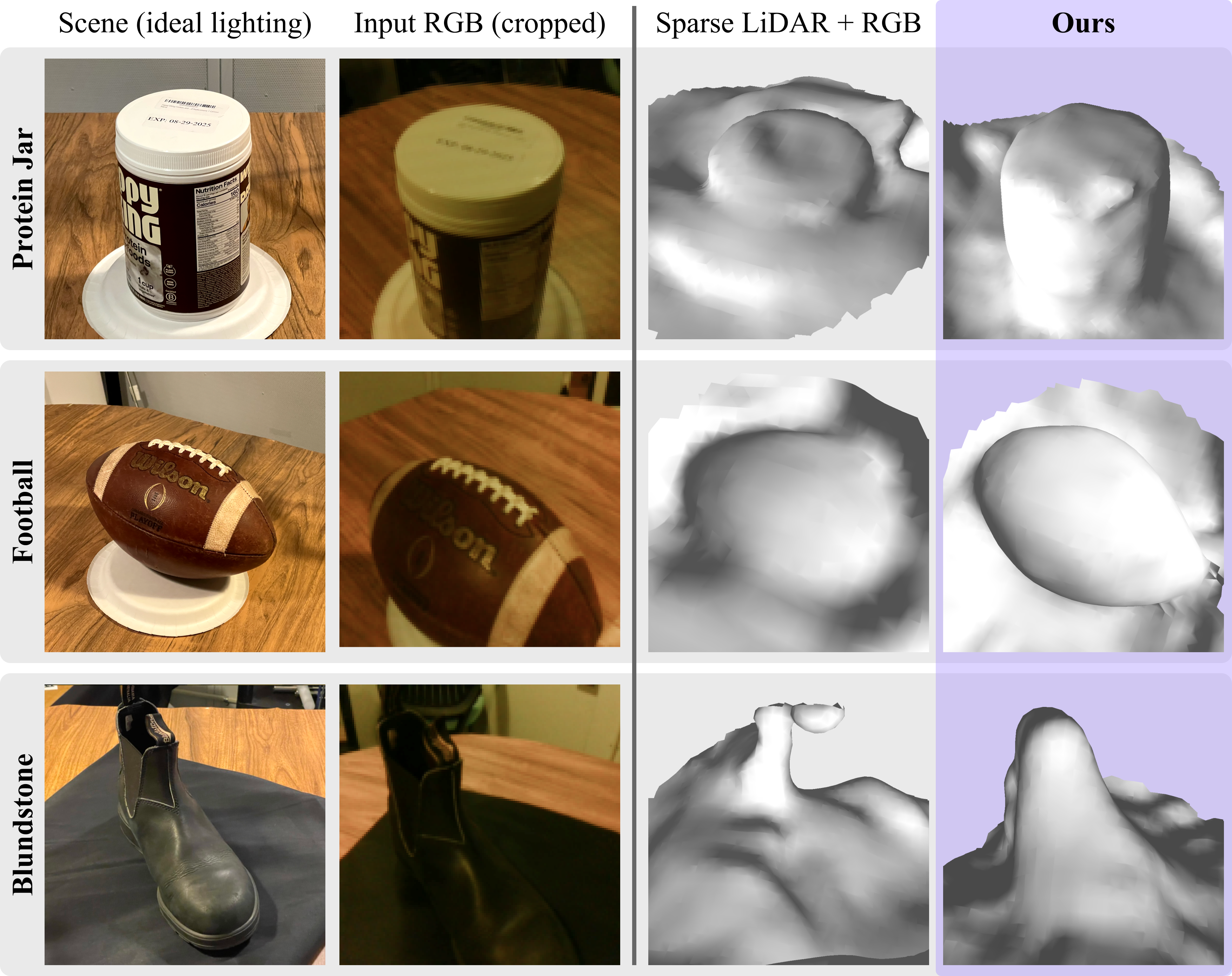}
\vspace{-6mm}
\caption{\textbf{Qualitative comparisons of real data captures.} We improve mesh reconstruction in challenging real-world scenes on few (90) inputs. RGB with sparse LiDAR fails to separate object and plane due to low albedo and poor spatial coverage, while our diffuse LiDAR improves boundary and geometry estimation. In very challenging settings (Blundstone), our method fails to separate object and plane, but nevertheless provides improved shape. }
\label{fig:comparison_real}
\vspace{-6mm} 
\end{figure}

\vspace{1mm}
\noindent
\textbf{Results.} We present quantitative and qualitative results from our simulated experiments for object texture variation in \cref{table:baseline-results} and \cref{fig:comparison_sim}. We find that our technique effectively combines diffuse LiDAR and RGB to enhance color and geometry compared to baselines, and in particular over RGB with sparse LiDAR. In full-texture scenes (\cref{table:baseline-results}a \& \cref{fig:comparison_sim}a), our scene-adaptive loss enables higher reliance on RGB, with diffuse LiDAR aiding depth estimation in peripheral regions with limited views (e.g. distant planes at oblique angles). In mixed-texture (\cref{table:baseline-results}b-c \& \cref{fig:comparison_sim}b-c) scenes, our scene-adaptive loss effectively prioritizes diffuse LiDAR in patches where RGB has less signal, yielding both improved ground plane separation and detail within textureless object convex hulls. In no-texture scenes (\cref{table:baseline-results}d \& \cref{fig:comparison_sim}d), we significantly outperform baselines, and enable sharper reconstructions and object-plane separation over RGB with sparse LiDAR.

We present simulated results for low-lighting robustness in \cref{fig:lowlighting}. We simulate low lighting with added Gaussian noise from $+\infty$ dB (no noise) to $-\infty$ dB (complete noise). At high SNR, our scene-adaptive loss can rely largely on RGB, resulting in only subtle benefits over baselines. As SNR decreases, we find our scene-adaptive loss can de-emphasize RGB and rely on diffuse LiDAR, enabling robust depth estimation even when noise fully obscures RGB cues. In these cases, RGB with sparse LiDAR struggles to reconstruct smooth geometry due to poor spatial coverage in LiDAR signals and the absence of multi-view RGB cues.

\subsection{Real-world Experiments}
\label{sec:real-experiments}

\textbf{Experimental Setup.} We capture diffuse LiDAR histograms using a low-cost AMS TMF8828~\cite{ams_osram_tmf882x} ToF sensor comprised of $18 \times 12$ SPAD pixels which are aggregated into $8 \times 8$ measurement zones on-device. The sensor is set to short-range, high-accuracy mode, increasing the temporal resolution but reducing the max measurement distance. RGB is captured with a co-located, rigidly mounted RealSense D435i module. We simulate a spot LiDAR by subsampling depth measurements from the D435i stereo IR depth map. We restrict input views to 90 captures uniformly sampled at a single elevation angle along the hemisphere. We describe this capture setup and sensor calibration in more detail in the supplement.


\begin{table}[!t]
\vspace{-0mm}
\centering
\caption{\textbf{LiDAR-only depth MAE ($\downarrow$) ablation.} Our reconstruction using diffuse LiDAR outperforms, on 10 training views, both sparse point LiDAR and sparse histogram LiDAR reconstruction.}
\vspace{-3mm}
\label{table:ablation_lidaronly}
\resizebox{0.99\linewidth}{!}{
\begin{tabular}{lcccc}
\toprule
\textbf{Method} & \textbf{Blender} & \textbf{Chair} & \textbf{Hotdog} & \textbf{Lego} \\
\midrule
Surfels - Sparse LiDAR Only & 0.078 & 0.099 & 0.084 & 0.074 \\
TransientNeRF \cite{malik2024transient} - Sparse Histograms & 0.281 & 0.221 & 0.222 & 0.228 \\
\textbf{Ours - Diffuse LiDAR Only} & \first 0.041 & \first 0.040 & \first 0.036 & \first 0.037 \\
\bottomrule
\end{tabular}
}
\vspace{-4mm}
\end{table}

\vspace{1mm}
\noindent
\textbf{Experimental Results.} We provide a qualitative comparison of our technique against an RGB and sparse LiDAR-guided reconstruction for scenes with challenging low-texture, low-lighting, and low-albedo conditions. We show results for these scenes in \cref{fig:comparison_real}. We find that our technique is effective in improving robustness of surface reconstruction across these challenging settings. As in synthetic experiments, we observe improved ground-plane separation, better geometry estimation for textureless concavities not visible from the RGB hull, and the ability to overcome complete failure cases of RGB methods in limited-view low-lighting conditions. We note that in very challenging conditions with very low lighting, albedo, and texture (``Blundstone"), we observe poor object-plane separation; yet, even in this case, we estimate improved geometry over RGB with sparse LiDAR.

\vspace{2mm}
\noindent
\textbf{Ablations.} We compare our diffuse LiDAR approach to existing sparse LiDAR-based techniques for \textit{LiDAR-only} reconstruction. We compare depth MAE using LiDAR-only loss for 1) Surfels-based sparse LiDAR, 2) TransientNeRF \cite{malik2024transient}, which uses sparse histograms instead of point depths, and 3) ours with diffuse LiDAR loss only. We find that our approach consistently outperforms these sparse LiDAR baselines (shown in \cref{table:ablation_lidaronly}). Importantly, we achieve depth MAE values close to those in No Texture reconstruction (\cref{fig:comparison_sim}d), suggesting that our scene-adaptive loss is effective in dynamically using LiDAR in low-texture scenes. We provide an additional ablation of our scene-adaptive loss in the supplement, where we demonstrate that removing this scene-adaptive loss degrades reconstruction quality. 

\vspace{-2mm}
\section{Conclusion}
\label{sec:conclusion}


In this work, we demonstrate how to leverage the complementary strengths of RGB and blurred LiDAR sensors for robust handheld 3D scanning in low-texture, low-light, and low-albedo environments. We show through recoverability analysis, quantitative evaluation, and qualitative real world experiments that diffuse LiDAR can, counterintuitively, enable more robust 3D scanning in these challenging settings. Future work in this direction could explore the role of diffuse LiDAR for joint ego pose and 3D estimation, and analyze robustness to other scenarios such as moving objects and challenging material types. The proposed reconstruction technique also holds potential for robust 3D scanning in other domains that require robust, mobile 3D scanning such as AR, VR, and robotics. We believe that this work can unlock the potential for combining these unconventional yet widely available sensors with RGB for robust 3D vision. 

\paragraph{Acknowledgements.} Nikhil Behari is supported by the NASA Space Technology Graduate Research Opportunity (NSTGRO) Fellowship. Aaron Young and Siddharth Somasundaram are funded by the National Science Foundation Graduate Research Fellowship Program (NSF GRFP). Tzofi Klinghoffer is supported by the National Defense Science and Engineering Graduate (NDSEG) Fellowship.

{
    \small
    \bibliographystyle{ieeenat_fullname}
    \bibliography{main}
}


\end{document}